# DETECTING EARTH-LIKE EXTRA-SOLAR PLANETS FROM ANTARCTICA BY GRAVITATIONAL MICROLENSING*

Philip Yock
Department of Physics
University of Auckland
Auckland
New Zealand.
Email: p.yock@auckland.ac.nz

ABSTRACT

Earth-like extra-solar planets may be detected with 1-2m class telescopes using the gravitational microlensing technique. The essential requirement is the ability to be able to carry out continuous observations of the galactic bulge. A telescope situated at Dome A or Dome C in Antarctica would be eminently suitable. Two possible observing strategies are described here. One employs a 1m visible (V and I passbands) telescope, the other a 2m wide-field near infrared telescope. Either telescope could allow a rough measurement of the abundance of Earth-like planets in the Milky Way to be made in a few years. Useful measurements could also be made on stellar atmospheres.

1. INTRODUCTION

Observations of extra-solar planets made during the last ten years yielded the most surprising result that a few per cent of stars possess 'hot Jupiters'. These are planets resembling Jupiter but located very close to their parent stars, with orbital radii < 0.1 AU. The unexpectedness of the result highlights the need for further exploration of planetary systems in the Galaxy. The result on hot Jupiters was obtained by the radial velocity technique, and confirmed by the transit technique. In both cases, ground-based telescopes were employed. Subsequent observations by the radial velocity technique have demonstrated its ability to detect Jupiter-like planets out to a few AU, and also 'hot Neptunes'.

Other techniques are required to detect Earth-like extra-solar planets. At present there are two possibilities. One will make use of the transit technique carried out from space. This is the goal of the Kepler space mission, which will be in orbit during the four-year period 2008 - 2012. Kepler has been designed to be sensitive to Earth-like planets orbiting solar-like stars out to a maximum orbital radius of about 1 AU. The

---

other technique that is presently available is gravitational microlensing. This can detect Earth-like planets orbiting K and M type main sequence stars at orbital radii of about 1.5 - 3 AU using 1m-class telescopes at mid-southern latitudes [1]. It is thus complementary in sensitivity to Kepler. It is possible to improve the sensitivity of the gravitational microlensing technique by using larger telescopes [2], a space telescope [3], or a telescope located in Antarctica. Here the last of these possibilities is discussed.

In the following sections, an introduction to the gravitational microlensing technique for detecting extra-solar planets is given. This is followed by a brief survey of results already obtained by microlensing on extra-solar planets and other topics. Two possible schemes for detecting extra-solar planets from Antarctica using the gravitational microlensing technique are then described. Finally, brief conclusions are given.

2. PLANET DETECTION BY GRAVITATIONAL MICROLENSING

As is well known, the gravitational microlensing process utilises the gravitational field of a 'lens' star to produce a magnified but distorted image of a background 'source' star. If the lens and source stars are perfectly aligned, the image is a ring - the Einstein ring. The angular radius $\theta_E$ of the ring is $(4GM_l(D_s-D_l)/D_sD_lc^2)^{1/2}$ where $M_l$ is the mass of the lens star, and $D_l$ and $D_s$ are the distances to the lens and source stars respectively. Also, the angular width of the ring is equal to the angular radius of the source star $\theta_s$, and the surface brightness of the ring is the same as that of the source star [4], as depicted schematically below.

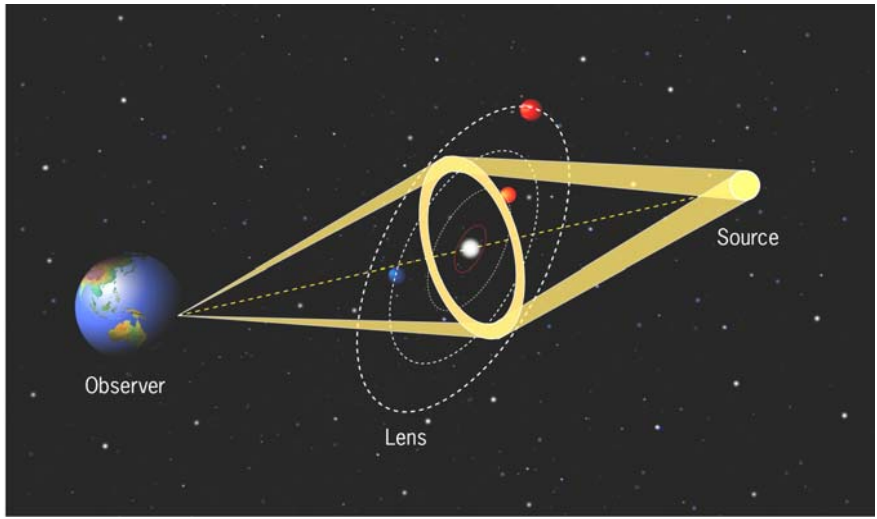

*Fig. 1.* The Einstein ring in a gravitational microlensing event with perfectly aligned lens and source stars. The radius of the ring is typically ~ 2-3 AU. Planets orbiting the lens star perturb the ring.

The above results imply that the magnification produced by a gravitational lens is simply $2\theta_E/\theta_s$. For solar-like lens and source stars located at ~ 4 kpc and ~ 8 kpc in the galactic disc and galactic bulge respectively, $\theta_E$ is very small, ~ 1 mas, but the magnification $A_{max}$ is very high, ~ 3000 [4]. As all stars are in motion, this magnification is only temporary. As the lens and source stars move into and out of alignment, the magnification first grows and then diminishes over a period of about a month, following a well known light-curve [4,5].

The absolute radius of the Einstein ring, $r_E = D_l\theta_E$, is typically about 2-3 AU. Thus, the Einstein ring coincides in size approximately to the orbits of asteroids in our Solar system. These occupy, of course, a central region in the Solar system. It is this coincidence that allows planets to be detected by gravitational microlensing. Planets orbiting a lens star may be situated rather close to the Einstein ring, as shown in Fig. 1, and may therefore deflect the lensed light to a detectable degree, even though they are much less massive than the central star of the lens. This is the essence of the gravitational microlensing technique for detecting planets.

In the vast majority of microlensing events, the lens and the source stars never come into perfect alignment. In these events the peak magnification $A_{max}$ is less than the value given above. It is given approximately by the smaller of either $2\theta_E/\theta_s$ or $\theta_E/u_{min}$. Here $u_{min}$ is the 'impact parameter' of an event, i.e. the minimum angular separation between the lens and source stars during an event. Typical values of $A_{max}$ range from a few (referred to as 'low magnification') to a few hundred (referred to as 'high magnification').

Also, most microlensing events involve either K or M type main-sequence stars as lenses, rather than solar-like stars. This does not change the numerical estimates given above greatly. It merely implies that most results on planets obtained by gravitational microlensing apply to extra-solar planets orbiting these types of stars, rather than solar-like stars. A large sample of microlensing events needs to be observed in order to obtain information on planets orbiting a broad range of host stars.

Initial attempts to detect extra-solar planets by gravitational microlensing were made in events of low magnification, simply because these events are more numerous. Planetary perturbations in low magnification events can be large [6-10], but they occur with low probability and at unpredictable times. Continuous monitoring of a large number of microlensing events is needed to detect them efficiently. This is difficult, but not impossible. Already, one detection was reported [11].

In contrast, planetary perturbations in events with high magnification occur with high probability and at predictable times [12]. Two-thirds of them occur within the FWHM of the light curves for these events, where the duration of the FWHM is typically about one day only [1]. For these reasons, systematic searches for planets can be mounted in high-magnification events relatively easily [13]. Already, two detections were reported [14-16].

The galactic bulge has proved to be a rich source of gravitational microlensing events, with several hundred events of low magnification being observed annually, together with at least ten events of high magnification. Two survey telescopes are presently being used to find these events. These are the OGLE 1.3m telescope in Chile [17], and the MOA 0.6m telescope in New Zealand [18]. Of these, the OGLE telescope provides the majority of detections [19,20]. However, beginning in 2006, the MOA group will commence observations with a new 1.8m survey telescope [18]. This should increase their detection rate.

Clearly, Antarctica offers the potential to monitor microlensing events found by the survey telescopes continuously, and hence the potential to detect planets, including Earth-like planets. Before proceeding to a discussion of how this could be achieved, we describe in the following section some results that have already been obtained by microlensing using telescopes at mid-southern latitudes.

3. SOME RESULTS OBTAINED BY MICROLENSING TO DATE

Microlensing studies have yielded results on planets, stars, dark matter, black holes, galactic structure and quasars, and also on the physics and theory of gravitational microlensing. Wambsganss reviewed most of these fields previously [21].

Studies of planets were carried out as described in Sect. 2. High-resolution studies of stars were carried out using gravitational lenses, especially binary lenses, to profile source stars. Studies of dark matter were the original goal of microlensers. In these studies, brown dwarfs or similar dark objects in the galactic halo ('MACHOs') were sought as lenses to stars in the Magellanic Clouds. Similar studies of lensed stars in the galactic bulge yielded information on galactic black holes, and on the bar-like structure of the galactic bulge. Quasar microlensing seeks to obtain high-resolution information on quasars by using galaxies with outlying stars as lenses. Information on the physics of microlensing has been obtained in several events where the geometry has been suitable.

Tables 1 and 2 below list some of the results that have been obtained to date on planets and stars by microlensing. These typify the results that could be expected to be obtained from Antarctica, as they rely on the ability to carry out continuous 'round-the-clock' observations of the galactic bulge during the southern winter. Fig. 2 illustrates some of the results included in the tables. The listed events do not comprise an exhaustive compilation of all events observed to date. In particular, some promising planetary events detected during 2005 that are still under analysis are not included.

| Event | Summary | Ref. |
|---|---|---|
| MACHO 98-BLG-35 | Evidence for Earth-mass extra-solar planet ($\Delta\chi^2 = 60$ only) | [14,15] |
| Various | Limit on abundance of Jupiters (<30%) | [22] |
| MACHO 99-LMC-2 | Search for extra-galactic planets | [15] |
| MOA-03-BLG-32/OGLE-03-BLG-192 | Sensitive search for Earth-mass extra-solar planets | [23,24] |
| OGLE-03-BLG-235/MOA-03-BLG-53 | Most distant planet (~ 5 kpc) | [11] |
| OGLE 04-BLG-343 | Highest magnification event (~ 3000) | [24] |
| OGLE 05-BLG-071 | Clearest planetary detection – Fig.2 (left panel) | [16] |

*Table 1. A sample of results obtained on planets by gravitational microlensing.*

| Event | Summary | Ref. |
|---|---|---|
| EROS-BLG-2000-5 | Limb darkening measurement of a K3III star | [25,26] |
| OGLE-02-BLG-069 | Profile of a G5III star in H alpha | [27] |
| MOA-02-BLG-33 | Limb darkening and shape measurement of solar-like star - Fig. 2 (right panel) | [28,29] |

*Table 2. A sample of results obtained on stars by gravitational microlensing.*

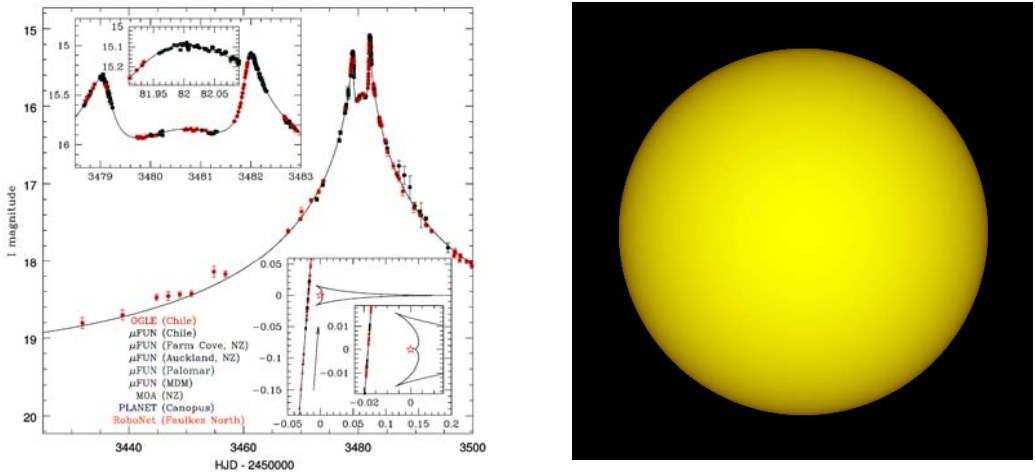

*Fig. 2. The left panel shows the light curve of microlensing event OGLE 2005-BLG-071 where the twin-peak structure was caused by an extra-solar Jupiter-like planet [16]. The right panel shows the image of solar-like star at a distance of 5 kpc obtained in event MOA 2002-BLG-33 [28,29]. For both events the observations were carried out with several telescopes at mid-southern latitudes. A single telescope located in Antarctica would have enabled seamless observations of the FWHMs of both events to be made, and hence would have enabled sensitive searches for Earth-mass planets to be carried out in both events.*

4. FOLLOW-UP TELESCOPE FOR OBSERVING THE PEAKS OF MICROLENSING EVENTS OF HIGH MAGNIFICATION FROM ANTARCTICA

Given that two survey telescopes are already operating whose function is to find gravitational microlensing events, especially those of high magnification, the obvious way to assist these programmes from Antarctica would be to operate a 'follow-up' telescope there to monitor the peaks of detected events. As remarked above, the peaks of these events contain the critical information.

A fairly modest telescope in Antarctica could do a superb job. This is because typical high-magnification events have solar-like source stars in the bulge with I ~ 20. When magnified by about 100, the magnitudes reach I ~ 15. This allows photometry to be carried out with a few tenths of a percent precision in an exposure time ~ 200 secs.

These are the requirements for detecting Earth-mass planets in these events [1]. For similar reasons, detailed observations of stellar atmospheres could be made using such a telescope, surpassing those listed in Table 2.

The extreme latitude of the Antarctic warrants special attention. Most gravitational microlensing events currently being found by the survey telescopes have declinations in the range -26° to -30°. During any 24-hour period, the zenith angles for these events would range from 51° - 73° for a telescope situated at Dome A at 81°S, and from 44° - 80° for a telescope at Dome C at 74°S. The largest of these zenith angles (80°) corresponds to 5.6 air masses [30]. This would present difficulties, in terms of atmospheric losses, differential refraction, and seeing.

By operating in the I-band, atmospheric losses would be limited to 0.3 mag [30]. This should be manageable using difference imaging [31]. Differential refraction would effectively introduce an additional 0.6 arcsec or less to the normal seeing [30]. Again, this could be handled using difference imaging. (It might be necessary to use separate reference images for difference imaging at zenith angles of 50°, 60°, 70° and 80°.) The additional seeing introduced by operating at large air masses is not yet known [32]. If it is comparable to, or not much greater than, the effective seeing introduced by differential refraction, it would also be manageable. Otherwise, an additional mid-southern follow-up telescope at the opposite longitude of the Antarctic telescope might be needed to obtain perfectly seamless data.

One final practical consideration is necessary. This concerns V-band images. For every event it is necessary to obtain colour information at high magnification in order to identify the source-star type. This could be achieved by imaging each event in the V-band when the bulge is highest in the sky. One or two exposures should suffice for this purpose.

The event rate for the above set-up would be determined by the survey telescopes. At present, about ten high magnification events that are suitable for planet hunting are being detected annually, and a similar number of events with binary lenses that are suitable for probing stellar atmospheres. These numbers might double when the new MOA survey telescope comes into operation in 2006. Every event that was monitored from Antarctica would be expected to yield significant information, either on extra-solar planets or on stellar atmospheres. Over the course of a few years, useful statistics on the abundance of low-mass planets orbiting low-mass stars would be obtained, and also useful information on the atmospheres of a variety of types of stars. These results are assured.

## 5. SURVEY TELESCOPE FOR MICROLENSING AT ANTARCTICA

A wide-field survey telescope dedicated to gravitational microlensing at Dome A or Dome C could achieve more than the set-up described above. It could detect planets in gravitational microlensing events of both high and low magnification, thereby increasing the detection rate. In addition, its spectral response would not need to be matched to those of existing survey telescopes. It could be operated in the near infrared (J+H) passband using HgCdTe detectors to take advantage of the exceptionally dry air above Antarctica [33]. It could then provide important information on lens stars that is not normally available when operating in visible passbands.

In typical microlensing events being detected by the existing survey telescopes that operate in visible passbands, the lens is a K or M dwarf and the source is an F or G dwarf. Before and after microlensing occurs, light from the lens is difficult or impossible to detect, unless one is willing to wait some years for the lens and the source stars to diverge from one another. During lensing, light from the lens star is even more difficult to detect. This facilitates observation of the microlensing event itself, but renders the identification of the lens star difficult. This is a significant deficiency, because one would like to have information not only on extra-solar planets, but also on the parent stars they are orbiting.

By operating a survey telescope in the J+H passband, microlensing events would be preferentially detected in which both the source and lens stars are K or M dwarfs. This would facilitate the detection of light from the lens star, and hence its identification. Moreover, the event rate would be increased, merely because lower mass stars are more numerous. In addition, one would be able to work closer to the centre of the Galaxy where the density of stars (both source and lens stars) is greater. If extinction was not too great, the event rate could be boosted.

Considerations similar to those discussed above for operating at large air masses would apply. However, one might be able to utilise the HgCdTe detectors to work in regions of the bulge (or even the Southern Milky Way) with declination < -30°. This would restrict the air mass to < 3 if the observations were carried out from Dome A. Effective seeing due to differential refraction would then be reduced to approximately 0.3 arcsecs [30], and atmospheric losses would be negligible. However, data on seeing at large zenith angles at Dome A would be required to quantify the overall performance of the telescope.

Colour information on all events would be needed, as for the follow-up telescope described above. This could be obtained by taking daily exposures of all fields in the V-band from a mid-southern site, at least for brighter events.

The aperture of a survey telescope operating in the J+H passband would probably need to be 2m or thereabouts, in order for it to be able to detect K and M dwarfs at distances ~ 8 kpc in short exposures when unmagnified. If the field-of-view was a few square degrees, some 10 square degrees of the bulge could be monitored at a sufficiently high frequency to detect planets in microlensing events of both high and low magnification with high efficiency.

A higher planetary detection rate than that achieved with the follow-up telescope described above would be achieved, but, without a good knowledge of the seeing at large zenith angles, it is impossible to quantify the rate. At best, one might hope to achieve results approaching those anticipated for the Microlensing Planet Finder (MPF) space mission [34]. For MPF, the projected planet discoveries are 66 terrestrial planets, 100 ice giants and 3300 gas giants assuming our Solar system is typical. With such capability, it should be possible to determine the abundances of all types of planets in the Galaxy. In addition, useful results would also be obtained on stellar atmospheres by either MPF of a survey telescope dedicated to microlensing at Antarctica. New results could also be anticipated on galactic structure by working closer to the galactic centre than was previously possible.

6. CONCLUSIONS

The Antarctic offers the exciting prospect of determining the abundance of Earth-like planets in the Milky Way with modest resources. One might hope that funding agencies would enthusiastically endorse an international effort to find other Earths from one end of our Earth using nature's lenses. Two plans have been described here, one involving a 1m follow-up telescope, the other a 2m wide-field survey telescope. Useful results from the follow-up telescope would be assured as soon as it was deployed. Experience gained with this telescope would assist detailed planning of a subsequent survey telescope.

ACKNOWLEDGEMENTS

The author thanks Michael Burton, John Hearnshaw, Carl Pennypacker, Will Saunders, Denis Sullivan and Don York for discussion.